\documentclass[pre,twocolumn,showpacs,superscriptaddress,floatfix]{revtex4}
\usepackage{graphicx,amssymb,amsmath,bbold}

\def\nbZ{{\mathchoice {\hbox{$\sf\textstyle Z\kern-0.4em Z$}}
{\hbox{$\sf\textstyle Z\kern-0.4em Z$}} {\hbox{$\sf\scriptstyle
Z\kern-0.3em Z$}}  {\hbox{$\sf\scriptscriptstyle Z\kern-0.2em Z$}}}}

\newlength{\ldag}
\newcommand{\F}{\frac}
\newcommand{\bra}{\langle}       \newcommand{\ket}{\rangle}

\newcommand {\vphi}{\varphi}

\newcommand {\vect}    [1]{{\boldsymbol #1}}

\DeclareMathOperator{\Tr}{Tr}
\DeclareMathOperator{\sign}{sign}

\DeclareMathOperator{\diag}{diag}

\newcommand{\cra}{{a^\dagger}}
\newcommand{\ana}{{a^{\phantom\dagger}\hspace{-\ldag}}}
\newcommand{\crb}{{b^\dagger}}
\newcommand{\anb}{{b^{\phantom\dagger}\hspace{-\ldag}}}

\newcommand{\tS}{\widetilde{S}}

\newcommand{\comment}[1]{}

\begin{document}
\title{Entanglement entropy in collective models}

\author{Julien Vidal}
\affiliation{Laboratoire de Physique Th\'eorique de la Mati\`ere Condens\'ee, CNRS UMR 7600,
Universit\'e Pierre et Marie Curie, 4 Place Jussieu, 75252 Paris Cedex 05, France}

\author{S\'ebastien Dusuel}
\affiliation{Lyc\'ee Louis Thuillier, 70 Boulevard de Saint Quentin, 80098 Amiens Cedex 3, France}

\author{Thomas Barthel}
\affiliation{Institute for Theoretical Physics C, RWTH Aachen, 52056 Aachen, Germany}

\begin{abstract}
We discuss the behavior of the entanglement entropy of the ground state in various collective systems. Results for general quadratic two-mode boson models are given, yielding the relation between quantum phase transitions of the system (signaled by a divergence of the entanglement entropy) and the excitation energies.
Such systems naturally arise when expanding collective spin Hamiltonians  at leading order via the Holstein-Primakoff mapping. In a second step, we analyze several such models (the Dicke model, the two-level BCS model, the Lieb-Mattis model and the Lipkin-Meshkov-Glick model) and investigate the properties of the entanglement entropy in the whole parameter range. 
We show that when the system contains gapless excitations the entanglement entropy of the ground state diverges with increasing system size. We derive and classify the scaling behaviors that can be met.
\end{abstract}

\pacs{03.65.Ud, 03.67.Mn, 21.10.Ev, 73.43.Nq}

\maketitle
%
%
%

%
%
\section{Introduction}
%
%
One of the most striking aspects of quantum mechanics is certainly the superposition principle which has no counterpart in classical physics. A direct consequence of this principle is the entanglement of physical states whose characterization has been a subject of intense research in the recent years.  
However,  the interest for the entropy in quantum systems has first emerged in black holes physics \cite{Bombelli86} pre-empting the recent favour in condensed matter physics.  
The relationship between entanglement and quantum phase transitions (QPTs) has especially drawn much attention since the original works on one-dimensional (1D) spin systems \cite{Osterloh02,Osborne02, Latorre03,Latorre04_1}. In these studies, it has been shown that entanglement measures are very sensitive to the presence of a critical point in the phase diagram. For example, in the 1D quantum Ising model in a transverse field, the entanglement entropy of the ground state has been shown to be finite everywhere except at the critical field where it diverges. This entropy, which is the central topic of this paper is defined as follows. Suppose that we split the degrees of freedom of the system under consideration into two subsystems $\mathcal{A}$ and $\mathcal{B}$. The entanglement entropy  $\mathcal{E}$ of any pure quantum state $\rho=|\psi\ket\bra\psi|$ with respect to this bipartition is defined as the von Neumann entropy of the reduced density matrix
%
%
\begin{equation}
  \mathcal{E}=-\mathrm{Tr}_\mathcal{A}\left(\rho_\mathcal{A}\ln\rho_\mathcal{A}\right)
            =-\mathrm{Tr}_\mathcal{B}\left(\rho_\mathcal{B}\ln\rho_\mathcal{B}\right),
    \label{eq:def_entropy}
\end{equation}
%
%
where $\rho_{\mathcal{A},\mathcal{B}}=\mathrm{Tr}_{\mathcal{B},\mathcal{A}} \: \rho$. Several fundamental questions arise concerning this entropy. In particular when, why  and how does it diverge? 
In $d$-dimensional non-critical (gapped) systems, the entanglement entropy is generically proportional to the surface area $L^{d-1}$ of the considered subsystem with linear size $L$ (\emph{area law}).
In critical systems, logarithmic corrections to this scaling may occur. In particular, a logarithmic scaling $\mathcal{E} \propto \ln L$ is found in critical one-dimensional boson and fermion (and thus also spin) systems \cite{Latorre03,Latorre04_1,Refael04,Peschel04,Its05,Keating04,Keating05,Its06,Santachiara06}. Note that in conformal field theory, the prefactor is determined by the central charges \cite{Korepin04,Calabrese04}. 
In higher-dimensional critical systems, the scaling behavior is different for bosons and fermions. For bosons, field theoretical and numerical results show that the area law prevails also in the critical case \cite{Bombelli86,Srednicki93,Callan94,Holzhey94,Barthel06_1}. For fermions, the area law is in this case corrected by a logarithmic factor \cite{Wolf06,Gioev06,Barthel06_1}. Note also that recently, noncritical fermionic systems with rapidly decaying interactions have been shown to violate this area law \cite{Eisert06}.

All the above cited studies focus on low-dimensional systems. The aim of the present work is to investigate collective systems which, in a sense, can be seen as infinite-dimensional. The physics of these models can be easily analyzed by a classical approach but the entanglement entropy comes from the quantum fluctuations around the classical ground state.
Here, we map those collective systems onto models with two interacting 
bosonic modes, each of them corresponding to one part of a bipartite splitting of the original system. We show that the scaling of the entanglement entropy depends mainly on two factors: the number of Goldstone modes (0, 
1 or 2) or, when approaching a critical point, the number of vanishing 
excitation gaps (1 or 2).
After giving the general results for quadratic two-mode boson Hamiltonians, we discuss in detail several collective models covering different behavior of entropy scaling.

The structure of this paper is the following.
In Sec. \ref{sec:general}, we derive the expression of the ground state entanglement entropy for general quadratic two-mode boson Hamiltonians and extract its explicit dependence on the excitation energies. Section \ref{sec:Dicke} is devoted to the study of the Dicke model for which the entropy is shown to diverge as $\F{1}{6} \ln N$ at the critical point where $N$ is the number of two-level systems. In Sec. \ref{sec:BCS}, we analyze the two-level BCS model in which the entropy diverges as 
$\F{1}{2} \ln N$ in the broken phase and as $\F{1}{3} \ln N$ at the critical point. In the antiferromagnetic long-range Lieb-Mattis model discussed in Sec. \ref{sec:Lieb}, we exactly compute the ground state entanglement entropy which behaves as $\ln N$. Section \ref{sec:LMG} focuses on the Lipkin-Meshkov-Glick (LMG) model for which we analyze the dependence of the entropy on several parameters (anisotropy, system and subsystem sizes). In each case, we compare our analytical predictions based on the scaling hypothesis (see Appendix \ref{App:scaling}) with numerics.

%
%
\section{General quadratic two-mode boson model}
\label{sec:general}
%
%

Let us consider the most general quadratic two-mode boson Hamiltonian
%
%
\begin{equation}
\label{eq:TwoBosonGen-HamVW}
 H=(\vect{b}^\dag)^T V \vect{b}+\F{1}{2} \Big[(\vect{b}^\dag)^T W \vect{b}^\dag+ {\rm h.c.} \Big] \,,
\end{equation}
%
%
where $\vect{b}=(b_1,b_2)$. We restrict here to the case where the coupling matrices $V$ and $W$ are real so that there are six free parameters. The aim is to find a simple expression for the entanglement entropy between both modes. As shown in Ref. \cite{Barthel06_1}, the reduced density matrix obtained by tracing over one of the two modes can be expressed in terms of the Green function matrices defined as $G^{++}_{ij}=\bra (b_i^\dag+b_i)(b_j^\dag+b_j) \ket$ and $G^{--}_{ij}=\bra (b_i^\dag-b_i)(b_j^\dag-b_j) \ket$. It is therefore convenient to reparametrize the Hamiltonian \eqref{eq:TwoBosonGen-HamVW} as  
%
%
\begin{subequations}
\label{eq:TwoBosonGen-HamVWparam}
\begin{gather}
V-W = R^\dag(\vphi)\,\omega\, R(\vphi)\,,\\
V+W =R^\dag(\vphi)\,\omega^{-\F{1}{2}}\, R^\dag(\psi)\,\Delta^2\, R(\psi)\,\omega^{-\F{1}{2}}R(\vphi)\,,
\end{gather}
\end{subequations}
%
%
where 
%
%
\begin{equation}
R(\vphi)=
\begin{bmatrix}
\cos\vphi &\sin\vphi\\
-\sin\vphi &\cos\vphi
\end{bmatrix}\, ,
\end{equation}
%
%
$\omega=\diag(\omega_+,\omega_-)$ and $\Delta=\diag(\Delta_+,\Delta_-)$. The matrices $V\pm W$ need to be positive definite, because the single-particle energies are necessarily positive. This is given, if we restrict to $\omega_\pm>0$. After a Bogoliubov transformation, $H$ is diagonalized in terms of new bosonic operators $\tilde{\vect{b}}$ \cite{Colpa78},
%
%
\begin{equation}
\label{eq:TwoBosonGen-HamL}
 H=(\tilde{\vect{b}}^\dag)^T \,\Delta\, \tilde{\vect{b}}+\F{1}{2}\Tr(\Delta-V)\,,
\end{equation}
%
%
and the Green's function matrices read
%
%
\begin{subequations}
\label{eq:TwoBosonGen-Green}
\begin{gather}
G^{++} = R^\dag(\vphi)\,\omega^{1/2}\, R^\dag(\psi)\,\Delta^{-1}\,R(\psi) \,\omega^{1/2}\,R(\vphi)\,,\\
G^{--}= - R^\dag(\vphi)\,\omega^{-1/2}\, R^\dag(\psi)\,\Delta \,R(\psi) \,\omega^{-1/2}\,R(\vphi)\,.
\end{gather}
\end{subequations}
%
%
The entanglement entropy between modes 1 and 2 is then given by \cite{Barthel06_1}
%
%
\begin{subequations}
\begin{gather}
\label{eq:Entropy_def}
\mathcal{E}=\F{\mu+1}{2}\ln\F{\mu+1}{2}-\F{\mu-1}{2}\ln\F{\mu-1}{2}\,,\\
\mu=\sqrt{-G^{++}_{1,1} G^{--}_{1,1}}=\sqrt{-G^{++}_{2,2} G^{--}_{2,2}}.
\label{eq:mudef}
\end{gather} 
\end{subequations}
%
%
Now, let us express all quantities in terms of the initial coupling matrices. Therefore, let us set:
%
%
\begin{equation}
V-W=
\begin{bmatrix}
d_0 &d\\
d &d_1
\end{bmatrix}\, ,
\end{equation}
%
%
which leads to 
%
%
\begin{equation}
\omega_\pm=\F{1}{2} \Big [d_0+d_1 \pm \epsilon \sqrt{4d^2+(d_0-d_1)^2} \Big] \,,\\
\end{equation}
%
%
where $\epsilon=\sign (d_0-d_1)$.

Similarly, setting: 
%
%
\begin{equation}
V+W=
\begin{bmatrix}
s_0 &s\\
s &s_1
\end{bmatrix}\, ,
\end{equation}
%
%
one obtains the excitation energies:
\begin{widetext}
%
%
\begin{equation}
\label{eq:gap_general}
\Delta_\pm=\sqrt{\F{1}{2} \Big [2ds+d_0 s_0+d_1 s_1 \pm \epsilon' \sqrt{(2ds+d_0 s_0+d_1 s_1)^2-4(d_0 d_1-d^2)(s_0 s_1-s^2)} \Big] }\,, 
\end{equation}
%
%
with 
%
%
\begin{equation}
\epsilon'=\sign \{\F{\epsilon 2d[s(d_0+d_1)+d(s_0+s_1)]+(d_0-d_1)(d_0s_0-d_1s_1)}
{\sqrt{4d^2+(d_0-d_1)^2}} \} \, .
\end{equation}
%
%
Of course, $\Delta_\pm$ have to be positive, giving the parameter range for which this transformation is well-defined. It is also useful to introduce the two parameters 
%
%
\begin{equation}
t=\tan(2\varphi)=2d/(d_0-d_1) \, ,
\end{equation}
%
%

%
%
\begin{equation}
u=\tan(2\psi) =\F{2(d_0 d_1-d^2)^{1/2} [s(d_0 -d_1)-d(s_0-s_1)]}
{2d[s(d_0+d_1)+d(s_0+s_1)]+(d_0-d_1)(d_0s_0-d_1s_1)} \, .
\end{equation}
%
%
\end{widetext}
The quantity $\mu$ defined in Eq. (\ref{eq:mudef}) is then given by
%
%
\begin{equation}
\label{eq:musimp}
\mu=\sqrt{1+(X_- \kappa +X_+  \kappa^{-1})^2}\, ,
\end{equation}
%
%
with
%
%
\begin{eqnarray}
X_\pm=\F{t (1+u^2)^{1/2} (\nu-\nu^{-1} ) \pm [2u+t(\nu+\nu^{-1} ) ]}
{4(1+t^2)^{1/2}(1+u^2)^{1/2}}\, ,
\end{eqnarray}
%
%
and 
%
%
\begin{equation}
\nu=\sqrt{\F{\omega_+}{\omega_-}}, \quad 
\kappa=\sqrt{\F{\Delta_+}{\Delta_-}}\, .
\end{equation}
%
%
Note that these expressions are still valid for the special case $d_0=d_1$ provided one sets $\epsilon=\sign(d)$.

This simple form allows one to investigate the various behaviors of the entropy that can be met in two-mode boson systems. Indeed, Eq. (\ref{eq:musimp}) clearly states that the entropy diverges when any of the quantities ${\omega_\pm,\Delta_\pm}$ vanishes or diverges and several cases must then be distinguished.

In the following, we will only focus on systems for which gaps vanish, signalling the presence of a quantum phase transition. In the next section, we consider the Dicke model \cite{Dicke54} in which one gap vanishes at the critical point but the other remains finite. In Sec. \ref{sec:BCS}, we consider the two-level BCS model for which both gaps vanish simultaneously at the transition. We show how these different behaviors lead to different finite-size scaling for the entropy at the critical point.
%
%
\section{The Dicke model}
\label{sec:Dicke}
%
%
Let us consider the single-mode Dicke Hamiltonian \cite{Dicke54} which describes the interaction of $N$ two-level systems with a single bosonic mode
%
%
\begin{equation}
   H=\omega_0 S_z+ \omega \cra \ana + {\lambda \over \sqrt{N}} \left( \cra + \ana \right) \left(S_+ + S_-\right)\, ,
   \label{eq:hamiltonian}
\end{equation}
%
%
where $\cra$ and $\ana$ are bosonic creation and annihilation operators satifying $[\ana,\cra]=1$. 
The collective spin operators are defined as  $S_{\alpha}=\sum_{i=1}^N \sigma_{\alpha}^{i}/2$ where the $\sigma_{\alpha}$'s are the Pauli matrices, and $S_\pm=S_x \pm {\rm i} S_y$. 

We refer the reader to Ref. \cite{Brandes05_1} for a description of the phase diagram as well as the super-radiance physics in this model. Here, we focus on the entanglement entropy studied in Refs. \cite{Lambert04,Lambert05} and discuss the linear entropy recently analyzed in Ref. \cite{Liberti06}. Let us also mention that other entanglement measures such as the concurrence have been studied in this model \cite{Lambert04,Reslen05_1,Vidal06_2}.

Our goal here is not to reproduce the results of Lambert et al. \cite{Lambert04} but to show how to extract the finite-size behavior of the entropy at the critical point.
Thus, we only consider the normal (symmetric) phase $(\lambda < \lambda_{\rm c}=\sqrt{\omega \omega_0}/2)$ characterized by  
$\lim_{N \rightarrow \infty} \langle \cra \ana \rangle/N=0$. The Holstein-Primakoff (HP) boson representation of the angular momentum \cite{Holstein40} reads:
%
%
\begin{eqnarray}
\label{eq:HP1}
S_+ &=&\crb N^{1/2}(1 - \crb \anb/N)^{1/2}=\left(S_-\right)^\dagger\, ,\\
\label{eq:HP2}
S_z &=& \crb \anb -{N \over 2}\, ,
\end{eqnarray}
%
%
with $[\anb,\crb]=1$, so that we now have to deal with a two-boson problem. 
In the thermodynamical limit and in the normal phase, one has $ \langle \crb \anb \rangle /N \ll 1$ and we can expand the square root in (\ref{eq:HP1}) to obtain the following form of the Hamiltonian:
%
%
\begin{eqnarray}
   H&=&-{N \over 2} \omega_0+\omega_0 \crb \anb+ \omega \cra \ana + \lambda \left( \cra+ \ana\right) \left( \crb+ \anb \right) \nonumber \\
&&
+\mathcal O\left(1/N \right)\, .
   \label{eq:hamiltonianD_expanded}
\end{eqnarray}
%
%

At this order, one thus has to deal with a quadratic Hamiltonian and we can make use of the results given in Sec. \ref{sec:general}. As found in Ref. \cite{Lambert05},  the two excitation energies are given by 
%
%
\begin{eqnarray}
\Delta_\pm&=&\frac{\sqrt{\omega ^2+\omega _0^2 \pm\sqrt{16 \lambda ^2\omega  \omega _0 +\left(\omega ^2-\omega_0
   ^2\right)^2}}}{\sqrt{2}}\, .
   \end{eqnarray}
%
%
Note that $\Delta_\pm$, as defined in Eq. (\ref{eq:gap_general}), depends on $\varepsilon=\sign(\omega ^2-\omega _0^2)$ which is assumed here to be positive without loss of generality.
At the critical point $\lambda_{\rm c}=\sqrt{\omega \omega_0}/2$, the system undergoes a second-order QPT and $\Delta_-$ vanishes as $(\lambda_{\rm c}-\lambda)^{1/2}$ whereas $\Delta_+=\sqrt{\omega^2+ \omega_0^2}$ remains finite. 

In this model, the parameter $\mu$ is given by:
%
%
\begin{equation}
\mu^{(0)}=\sqrt{1+\F{4 \omega \omega_0 \lambda^2}{16 \lambda ^2\omega  \omega _0 +\left(\omega ^2-\omega_0^2\right)^2}
   \bigg(\F{\Delta_+}{\Delta_-}+\F{\Delta_-}{\Delta_+} -2\bigg)}  \, ,\nonumber\\
\end{equation}
%
%
where the superscript "$(0)$" refers to the order in the $1/N$ expansion. The entanglement entropy between modes $a$ and $b$ is then given by Eq. (\ref{eq:Entropy_def}).
In the vicinity of the critical point, the entropy thus diverges as:
%
\begin{equation}
\label{eq:entropyDDL0}
\mathcal{E}^{(0)}= 1+\ln \left[\frac{\lambda_{\rm c}^5}{2(\lambda_{\rm c}-\lambda)\left(\omega ^2+\omega _0^2\right)^2}\right]^{1/4}+ \mathcal O[(\lambda_{\rm c}-\lambda)^{1/2}]\, .
\end{equation}
%
We note that this expression is in agreement with those obtained in Ref. \cite{Lambert05} with a different approach. However, the constant term in this latter study is wrong and we give here the correct value.

To extract the behavior of the entropy at the critical point, one simply invokes the scaling hypothesis discussed in Ref. \cite{Vidal06_2} for this model and detailed in Appendix \ref{App:scaling}. In the present context, one indeed has, near the critical point, $\mathcal{E}^{(0)} \sim -\F{1}{4} \ln (\lambda_{\rm c}-\lambda)$ 
which yields $\mathcal{E} \sim \F{1}{6} \ln N$. Note that this exponent $1/6$ is close to the numerical result  $(0.14 \pm 0.01)$  obtained from exact diagonalizations \cite{Lambert05}.

Similarly, one can easily compute the linear entropy which reads :
%
%
\begin{equation}
  \label{eq:entropylinD0}
\mathcal{E}^{(0)}_{\rm lin.}=1-\frac{1}{\mu^{(0)}}\, ,
\end{equation}
%
%
as shown in Appendix \ref{App:entropy_lin}.  In the vicinity of $\lambda_{\rm c}$, one has
%
\begin{equation}
\label{eq:entropylinDDL0}
\mathcal{E}^{(0)}_{\rm lin.} =1-\left[\frac{(\lambda_{\rm c}-\lambda)\left(\omega ^2+\omega _0^2\right)^{2}}{8\lambda_{\rm c}^{5}}\right]^{1/4}+ \mathcal O[(\lambda_{\rm c}-\lambda)^{3/4}]\, .
\end{equation}
%
Consequently, using the scaling hypothesis, and since 
$\mathcal{E}^{(0)}_{\rm lin.} -1\sim (\lambda_{\rm c}-\lambda)^{1/4} $
one expects $\mathcal{E}_{\rm lin.}-1 \sim N^{-1/6}$ at the transition as predicted in Ref. \cite{Liberti06} in the adiabatic regime $(\omega/\omega_0 \ll 1)$. Here again, we wish to emphasize that the expression given by Liberti {\it et al.} is only valid in this regime. Thus, their expression of the linear entropy is only correct at first order in $\omega / \omega_0$ as can be easily checked by expanding (\ref{eq:entropylinDDL0}) in powers of this parameter $\omega / \omega_0$.

To conclude this section, we mention that the same calculation can be performed in the broken phase $\lambda>\lambda_{\rm c}$ (see Ref. \cite{Lambert05}). 
Further, the exact correction of order $1/N$ could, in principle, be easily calculated but its expression would be too long to be given for this model. We shall compute it for the systems considered thereafter.
%
%
\section{The two-level BCS model}
\label{sec:BCS}
%
%
We now turn to another problem which can also be mapped onto a two-mode boson  system, but where both gaps vanish at the critical point. The two-level BCS model originates from superconductivity \cite{Bardeen57,vonDelft01} but can be translated into a simple spin model (see for instance Ref. \cite{Roman02}). Its entanglement properties have recently been discussed but only from the concurrence point of view \cite{Dusuel05_1,Dunning05}. 
Here, we shall show that its entropy captures some nontrivial features due to the vanishing gaps.  

The two-level BCS Hamiltonian reads
%
%
\begin{equation}
  \label{eq:ham_bcs_spins}
  H=-\frac{1}{N}\left( S_x^2+S_y^2\right) - h(S_z^{\mathcal{A}}-S_z^{\mathcal{B}}) \, .
\end{equation}
%
%
Here, the indices $\mathcal{A}$ and $\mathcal{B}$ refer to two distinct sets of $N/2$ spins $1/2$ subjected to antiparallel magnetic fields.  Here, we have introduced the collective spin operators 
$S_\alpha^{\mathcal{A},\mathcal{B}}=\sum_{i\in \mathcal{A},\mathcal{B}} \sigma_\alpha^i/2$, and $S_\alpha=S_\alpha^{\mathcal{A}}+S_\alpha^{\mathcal{B}}$ (where $\alpha=x,y,z$).
We further restrict our study to the total magnetization $S_z=0$ which is reminiscent of the fermion number conservation in the initial BCS problem.
Let us first discuss the symmetric phase $h>1$ characterized by 
$\lim_{N \rightarrow \infty} 4 \langle S_z^{\mathcal{A}} \rangle/N=-\lim_{N \rightarrow \infty} 4 \langle S_z^{\mathcal{B}} \rangle/N=1$. 
As previously, we perform a $1/N$ expansion using the HP boson representation of the spin operators:
%
\begin{eqnarray}
  S_z^{\mathcal{A}} &=& N/4-\cra \ana, \quad S_z^{\mathcal{B}} = \crb \anb -N/4\, ,\\
  S_+^{\mathcal{A}} &=& (S_-^{\mathcal{A}})^\dagger=(N/2)^{1/2}\left(1-2\cra \ana/N\right)^{1/2} \ana\, ,\\
  S_+^{\mathcal{B}} &=& (S_-^{\mathcal{B}})^\dagger=(N/2)^{1/2}\crb \left(1-2 \crb \anb/N\right)^{1/2}\, ,
  \end{eqnarray}
%
with $S_\pm^{\mathcal{A},\mathcal{B}}=S_x^{\mathcal{A},\mathcal{B}} \pm {\rm i} S_y^{\mathcal{A},\mathcal{B}}$ and $[\ana,\cra]=[\anb,\crb]=1$. 
In the thermodynamical limit and in the normal phase, one has $ \langle \cra \ana \rangle /N \ll 1$ and $\langle \crb \anb \rangle /N \ll 1$ so that the Hamiltonian can be written as:
%
%
\begin{eqnarray}
   H&=&-{N h \over 2} -\frac{1}{2}+ \Big(h-\frac{1}{2} \Big) (\cra \ana+\crb \anb) -\frac{1}{2}(\cra \crb+\ana \anb) \nonumber  \\
   && +\frac{1}{N}\Big(\cra \ana+\crb \anb+\cra^2 \ana^2+\crb^2 \anb^2\Big) \nonumber \\
   &&+\frac{1}{2 N}\Big(\ana \crb \anb^2+\cra \crb^2 \anb+ \cra \ana^2 \anb +\cra^2 \ana \crb\Big) \nonumber \\
   &&+ \mathcal O (1/N^2)\, .
   \label{eq:hamiltonianBCS_expanded}
\end{eqnarray}
%
%
At order $N^0$, the Hamiltonian is easily diagonalized and one then finds that, in this phase,  both gaps are equal to $\Delta=\sqrt{h(h-1)}$. Thus, at the critical point $h=1$, both gaps vanish $(\kappa=1)$.  

The parameter $\mu$ can however still be expressed in terms of the gap 
%
%
\begin{equation}
\mu^{(0)}=\frac{2h-1}{2 \Delta}\, ,
\end{equation}
%
%
where the superscript "$(0)$" again refers to the order in the $1/N$ expansion.
The entanglement entropy between sets $\mathcal{A}$ and $\mathcal{B}$, in the thermodynamical limit and in the symmetric phase is then given by Eq. (\ref{eq:Entropy_def}). In the vicinity of the critical point, the entropy thus behaves as:
%
\begin{equation}
\mathcal{E}^{(0)} = 1-2 \ln 2-\ln (h-1)^{1/2} +\mathcal O(h-1)\, .
\end{equation}
%
The scaling hypothesis (see Appendix \ref{App:scaling}) immediately implies $\mathcal{E} \sim \F{1}{3} \ln N$ at the transition point.

\begin{figure}[t]
  \centering
\includegraphics[width=8cm]{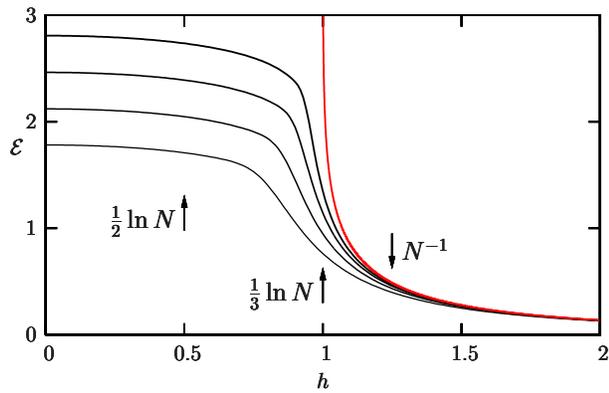}
  \caption{(Color online) Entanglement entropy in the two-level BCS model  as a function of $h$ for $N=32,64,128,256$ (from numerics [black lines]) and $\infty$ ($\mathcal{E}^{(0)}$ [red line]). Arrows indicate the behavior of the finite-size correction in various regions. }
  \label{fig:BCS_sym}
\end{figure}

\begin{figure}[t]
  \centering
\includegraphics[width=8cm]{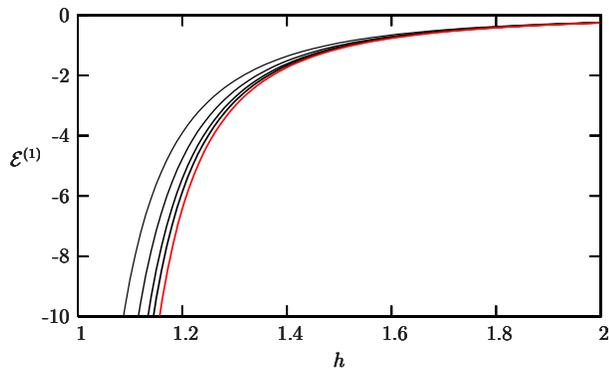}
  \caption{(Color online) Behavior of $N\big[\mathcal{E}_{\rm num}-\mathcal{E}^{(0)}\big]$ as a function of $h$ in the two-level BCS model for  $N=32,64,128,256$ (from numerics [black lines]) and $\infty$ ($\mathcal{E}^{(1)}$ [red line]). }
  \label{fig:BCS_correction}
\end{figure}

Before comparing with numerics and discussing the broken phase, we also give the next order correction which can be computed exactly in this model. 
At order $1/N$, the Hamiltonian can be diagonalized using the canonical transformations method  \cite{Vidal06_2}. This allows one to evaluate the correction of order $1/N$ to $\mu$ 
%
%
\begin{equation}
\mu^{(1)}=\frac{h}{2} \bigg(- \frac{h}{\Delta^4}+\frac{1}{\Delta^{3}} \bigg) \, .
\end{equation}
%
%
As explained in Ref. \cite{Barthel06_4}, $\mu^{(1)}$ is sufficient to determine the first correction to the entropy (see Appendix \ref{App:correction}). The entropy at order $1/N$ is indeed obtained by Taylor expanding 
expression (\ref{eq:Entropy_def}) where $\mu$ is also computed at this order as above. 
One then obtains:
%
%
\begin{equation}
\mathcal{E}^{(1)} =\frac{ \mu^{(1)}}{2} \ln\bigg(\F{\mu^{(0)} +1}{\mu^{(0)} -1} \bigg)\, . 
\end{equation}
%
%
 This expression allows to check explicitly the scaling hypothesis (see Appendix \ref{App:scaling}) since, at this order, and in the vicinity of the critical point, one has
%
%
\begin{eqnarray}
{\rm e}^\mathcal{E}&=& {\rm e}^{\mathcal{E}^{(0)}+\frac{1}{N}\mathcal{E}^{(1)}} \,,\\
&\sim& (h-1)^{-1/2} \bigg[1-\frac{1}{N (h-1)^{3/2}} \bigg]\, .
\end{eqnarray}
%
%
As can be seen in Fig.  \ref{fig:BCS_sym},  an excellent agreement is found between the entanglement entropy for increasing $N$ and the thermodynamic limit. The leading finite $N$ correction is also extracted from numerics and compared to the analytical result (see Fig. \ref{fig:BCS_correction}) given above confirming the validity of the scaling hypothesis.
%
%
%
%

%
%
%
%

Let us now turn to the broken phase $h<1$. In this phase, at order $N^0$, one of the two gaps vanishes whereas the other remains finite. More precisely, one has
%
%
\begin{eqnarray}
\Delta_+&=&\sqrt{1-h^2}  \,,\\
\Delta_-&=&0 \, .
\end{eqnarray}
%
%
We do not give here the details of this analysis since it is similar to the isotropic LMG model discussed in the next section. 
Unfortunately,  one can not use the scaling hypothesis to determine the behavior of the entropy due to the presence of the Goldstone mode in this parameter range ($\Delta_-=0$). However, one can exactly determine the entropy in zero field. There, the ground state is given by the Dicke state corresponding to $S= N/2$ and $S_z=0$ whose entropy is known \cite{Popkov05,Latorre05_2} to diverge as $\F{1}{2} \ln N$. Since no drastic physical changes occur when varying the field below $h<1$, it is reasonable to expect a similar behavior in the entire broken phase. 
%
%
\begin{figure}[t]
  \centering
   \includegraphics[width= 8cm]{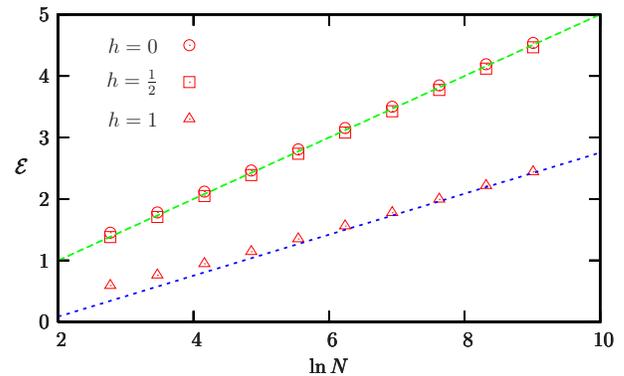}
  \caption{(Color online)
    Entanglement entropy as a function of $\ln N$ in the two-level BCS model for $h= 0, 1/2, 1$ and analytic predictions: slope $1/2$ (green line) for $h<1$, slope $1/3$ (blue line) for $h=1$.}
  \label{fig:BCS_broken}
\end{figure}
%
%
We have checked this handwaving argument with numerics and, indeed, the ground state entropy diverges as $\F{1}{2} \ln N$ in the entire broken phase (see Fig. \ref{fig:BCS_broken}). 

To complete this study, it is interesting to analyze a similar model for which the classical analysis predicts two-Goldstone modes instead of one. The Lieb-Mattis Hamiltonian \cite{Lieb62} is a natural candidate.
%
%
\section{The Lieb-Mattis model}
\label{sec:Lieb}
%
%
The Lieb-Mattis Hamiltonian \cite{Lieb62} is given by 
%
%
\begin{equation}
  \label{eq:ham_LM}
  H=\frac{1}{N} \: {\bf S^\mathcal{A}} \cdot {\bf S^\mathcal{B}}=\frac{1}{2N}
  \Big[{\bf S}^2-\big({\bf S^\mathcal{A}}\big)^2-\big({\bf S^\mathcal{B}}\big)^2\Big]\, .
\end{equation}
%
%
As for the two-level BCS model,  the indices $\mathcal{A}$ and $\mathcal{B}$ refer to two distinct sets of $N/2$ spins $1/2$. The (nondegenerate) ground state $|\psi_0 \ket$ is readily found to be the eigenstate of  $\Big\{{\bf S^2},\big({\bf S^\mathcal{A}}\big)^2,\big({\bf S^\mathcal{B}}\big)^2 \Big\}$, with eigenvalues $\Big\{0,\tfrac{N}{4} \big(\tfrac{N}{4}+1\big),\tfrac{N}{4} \big(\tfrac{N}{4}+1\big)\Big\}$. However, the classical ground state is found to have two Goldstone modes stemming from the isotropy of the Hamiltonian (\ref{eq:ham_LM}). This suggests that the scaling behavior of the entanglement entropy here should differ from those of the models considered previously. As in the broken phase of the BCS model, the diagonalization of the quadratic Hamiltonian that would derive from a HP expansion does not give the correct ground state because of the Goldstone modes. Nevertheless,  the entanglement entropy of the ground state can be evaluated exactly since:
%
%
\begin{equation}
  \label{eq:psi_0}
  |\psi_0 \ket= \F{1}{\sqrt{\F{N}{2}+1}} \sum_{M=-N/4}^{N/4} (-1)^{-M+N/4} | M,-M \ket \,,
\end{equation}
%
%
where the state $ | M,-M \ket$ denotes the eigenstate of 
$\Big\{ \big({\bf S^\mathcal{A}}\big)^2,S_z^{\mathcal{A}},\big({\bf S^\mathcal{B}}\big)^2,S_z^{\mathcal{B}} \Big\}$ with eigenvalues 
$\Big\{\tfrac{N}{4} \big( \tfrac{N}{4}+1 \big),M$, $\tfrac{N}{4} \big( \tfrac{N}{4}+1 \big), M \Big\}$. The reduced density matrix $\rho_\mathcal{A}$ obtained by tracing over set $\mathcal{B}$ is thus easily computed and reads in the eigenbasis of $ \Big\{ \big({\bf S^\mathcal{A}}\big)^2,S_z^{\mathcal{A}} \Big\}$
%
%
\begin{equation}
  \rho_\mathcal{A}= \Tr_{\mathcal{B}} |\psi_0 \ket  \bra \psi_0|=  \F{1}{\F{N}{2}+1} | M \ket \bra M | \, .
\end{equation}
%
%
The entanglement entropy is thus given by 
%
%
\begin{equation}
{\mathcal E}= -\Tr \rho \ln \rho=\ln\big(\tfrac{N}{2}+1\big)\, ,
\end{equation}
%
%
and thus behaves as $\ln N$ at large $N$.

To conclude this brief review of entanglement entropy in collective systems, we shall now consider a collective spin model with an arbitrary bipartition.

%
%
\section{The Lipkin-Meshkov-Glick model}
\label{sec:LMG}
%
%
Initially proposed fourty years ago by Lipkin, Meshkov and Glick \cite{Lipkin65,Meshkov65,Glick65} to describe phase transitions in nuclei, the LMG model has, since then, been widely used to describe many physical systems such as Bose-Einstein condensates \cite{Cirac98}, Josephson junctions or long-range interacting spin systems \cite{Botet82,Botet83}. Here, we shall consider the latter formulation which is especially well-suited to our problematics. The entanglement properties of the LMG model have drawn much attention in the last years \cite{Vidal04_1,Vidal04_2,Vidal04_3,Dusuel04_3,Dusuel05_2,Reslen05_1} but the entropy has been analytically investigated only very recently \cite{Barthel06_2} following the numerical work of Latorre {\it et al.} \cite{Latorre05_2}. 

The Hamiltonian of the LMG model is given by
%
%
\begin{equation}
  H=-\frac{1}{N} \big(S_x^2 + \gamma S_y^2\big) - h \: S_z\, ,
    \label{eq:ham_LMG}
\end{equation}
%
%
where $S_{\alpha}=\sum_{i} \sigma_{\alpha}^{i}/2$, and the $\sigma_{\alpha}$'s are the Pauli matrices. It describes a system of $N$ spin $1/2$ mutually interacting in a field $h$, transverse to the interaction directions ($XY$). Here, we restrict our study to the most interesting ferromagnetic case and, without loss of generality, we assume $0 \leqslant \gamma < 1$ and $h \geqslant 0$. The isotropic case $\gamma=1$ is discussed separately at the end of the section.

As was early discussed in the original work of Lipkin, Meshkov and Glick, this system undergoes a second-order QPT at $h=1$, between a symmetric ($h>1$) and a broken ($h<1$) phase. In both regimes, the ground state lies in the maximum spin sector $S=N/2$. To analyze the entanglement entropy, we partition the set of $N$ spins into two blocks $\mathcal{A}$ and $\mathcal{B}$, of sizes $L$ and $(N-L)$ respectively \cite{Barthel06_2}.  We also introduce the corresponding spin operators $S_{\alpha}^{\mathcal{A},\mathcal{B}}=\sum_{i \in {\mathcal{A},\mathcal{B}}} \sigma_{\alpha}^{i}/2$. 

\begin{figure}[t]
  \centering
  \includegraphics[width= 8cm]{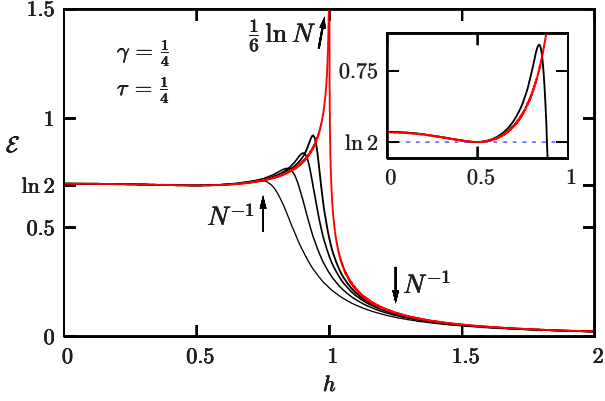}
  \caption{(Color online)
    Entropy as a function of $h$ in the LMG model for  $\gamma=1/4$, $\tau=1/4$ and $N= 32, 64, 128, 256$ (from numerics [black lines]) and $\infty$ ($\mathcal{E}^{(0)}$ red line). Arrows indicate the behavior of the finite-size correction in various regions. 
 For $h<1$, we plotted $\mathcal{E}^{(0)}+\ln 2$ to take into account the parity-broken nature of the classical states taken as starting point of the $1/N$ expansion \cite{Lambert05,Barthel06_2}. The inset is a zoom around $h=\sqrt{\gamma}$ where $\mathcal{E}^{(0)}=0$, for $N=64$ (black line) and $\infty$ (red line).}
  \label{fig:entropy_LMG}
\end{figure}

Let us first focus on the symmetric phase $h>1$ in which the classical ground state is fully polarized in the $z$-direction. As previously, it is convenient to use the HP representation of the spin operators
%
%
       \begin{eqnarray}
         S_z^\mathcal{A}&=&L/2 - \cra \ana \, , \label{eq:HP1A} \\
         S_-^\mathcal{A}&=&L^{1/2}  \: \cra  \:(1- \cra \ana/L)^{1/2}=(S_+^\mathcal{A})^\dag\, ,
         \label{eq:HP2A} \\
         S_z^\mathcal{B}&=&(N-L)/2 - \crb \anb \, , \label{eq:HP1B} \\
         S_-^\mathcal{B}&=&(N-L)^{1/2} \: \crb \:(1-\crb \anb/ (N-L)^{1/2}=(S_+^\mathcal{B})^\dag\, , \quad \quad
         \label{eq:HP2B}
       \end{eqnarray}
%
%
with $S_\pm^{\mathcal{A},\mathcal{B}}=S_x^{\mathcal{A},\mathcal{B}} \pm {\rm i} S_y^{\mathcal{A},\mathcal{B}}$. 
This transformation maps the LMG Hamiltonian onto a system of two bosonic modes. At fixed $\tau=L/N$, the Hamiltonian can be expanded in powers of $1/N$ since, in this phase, one has $\bra \cra \ana \ket /L \ll 1$ and $\bra \crb \anb \ket /(N-L) \ll 1$. 
At order $(1/N)$ and for $h > 1$, one then gets
\begin{widetext}
%
%
\begin{eqnarray}
  \label{eq:HP_LMG}
  H&=&-\frac{N h}{2}-\frac{1+\gamma}{4}+\frac{2h-\gamma-1}{2} \Big( \cra \ana+\crb \anb \Big)  
+\frac{\gamma-1}{4}  \Big[ \tau \Big( {\cra}^2+\ana^2 \Big) + (1-\tau) \Big( {\crb}^2+\anb^2 \Big) 
+2 \sqrt{\tau(1-\tau)} \Big( \cra \crb +\ana \anb \Big) \Big]  + \nonumber \\
&&\frac{1}{N} \bigg\{  
\F{1-\gamma}{4 \sqrt{\tau(1-\tau)}} 
\Big[(1-\tau) \Big( \cra \ana^2 \anb + {\cra}^2 \ana \crb \Big) +
\tau \Big( \ana \crb \anb^2 + {\cra} {\crb}^2 \anb \Big) \Big]
+ 
\frac{1+\gamma}{2}\Big( \cra \ana+\crb \anb +{\cra}^2 \ana^2 + {\crb}^2 \anb^2 +2  \cra \ana \crb \anb \Big) +\nonumber \\
&& \frac{1-\gamma}{8 } \Big[ {\cra}^2+\ana^2 + {\crb}^2+\anb^2 +2 \Big({\cra} \ana^3+{\cra} ^3 \ana +{\crb} \anb^3+{\crb} ^3 \anb \Big) \Big] \bigg\} +{\mathcal O}(1/N^2)  \, .
\end{eqnarray}
%
\end{widetext}
At order $N^0$, this Hamiltonian is easily diagonalized and the excitation energies are given by:
%
%
\begin{eqnarray}
\Delta_+&=&\frac{2h-\gamma-1}{2} \, ,\\
\Delta_-&=&\sqrt{(h-1)(h-\gamma)}\, .
\end{eqnarray}
%
%
Thus, at the critical point, $\Delta_-$ vanishes whereas $\Delta_+$ remains finite except for $\gamma=1$ (see discussion at the end of the section).
The parameter $\mu$ is then given by:
%
%
\begin{equation}
\mu^{(0)}=\sqrt{1+\tau(1-\tau) \Bigg[ \bigg(\F{h-1}{h-\gamma}\bigg)^{1/4} -
\bigg(\F{h-\gamma}{h-1}\bigg)^{1/4} \Bigg]^2}\, ,
\label{eq:LMG_mu0}
\end{equation}
%
%
which is equivalent to the form presented in Ref. \cite{Barthel06_2}. The entropy is then given by 
Eq. (\ref{eq:Entropy_def}). Figure \ref{fig:entropy_LMG} shows a comparison between the entropy at finite $N$ obtained from numerical diagonalization of the Hamiltonian and the entropy $\mathcal{E}^{(0)}$ obtained from Eqs. (\ref{eq:Entropy_def}) and (\ref{eq:LMG_mu0}). As can be seen, the entropy is finite for all $h \neq 1$ and diverges at the transition point. 

%
%
%
%

\begin{figure}[t]
  \centering
\includegraphics[width=8cm]{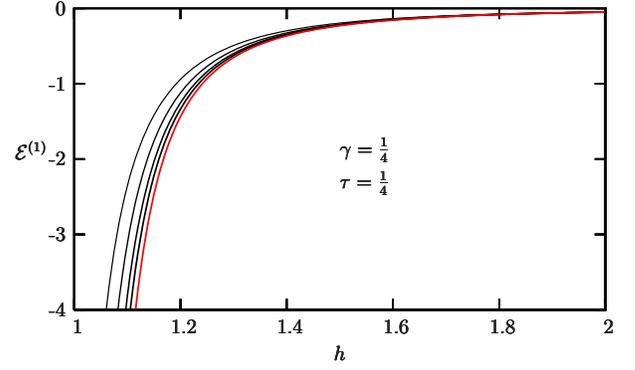}
  \caption{(Color online) Behavior of $N\big[\mathcal{E}_{\rm num}-\mathcal{E}^{(0)}\big]$ as a function of $h$ in the LMG model for  $N=32,64,128,256$ (from numerics [black lines]) and $\infty$ ($\mathcal{E}^{(1)}$ red line). }
  \label{fig:correction_LMG}
\end{figure}

Near the critical point $h=1^+$, it behaves as
%
\begin{eqnarray}
\label{eq:E0_LMG}
\mathcal{E}^{(0)}&=&{-\frac{1}{4} \ln (h-1)+ \frac{1}{2}\ln [\tau(1-\tau)]+ \frac{1}{4}\ln (1-\gamma) } +
\nonumber \\
&&1-\ln 2+\mathcal O[(h-1)^{1/2}]\, ,
\end{eqnarray}
%
where the superscript "$(0)$" refers to the order in the $1/N$ expansion. As in the Dicke model studied in Sec. \ref{sec:Dicke}, the scaling hypothesis (see Appendix \ref{App:scaling}) yields $\mathcal{E} \sim \F{1}{6} \ln N$. As can be seen in Eq. (\ref{eq:E0_LMG}), the entropy also diverges when $\gamma$ goes to 1 and when 
$\tau$ reaches its extremal values 0 or 1. This suggests a possible dependence of the scaling variable on these parameters. To obtain this dependence, one needs, at least,  to compute the correction of order $1/N$ to the entropy. 
As explained in Ref. \cite{Barthel06_4}, it is obtained by Taylor expanding expression (\ref{eq:Entropy_def}) at order $1/N$ where $\mu$ is also computed at this order (see Appendix \ref{App:correction}). This latter step requires to diagonalize $H$ at order $1/N$. Using the canonical tranformation method described in Ref. \cite{Vidal06_2}, one then obtains the correction of order $1/N$ to $\mu$
%
%
\begin{equation}
\mu^{(1)}=\F{\tau(1-\tau)(1-\gamma)^2}{2}\bigg[ \F{4 \gamma-h(1+2h+\gamma)}{2 \Delta_-^4}+\F{h}{\Delta_-^3}\bigg]\, .
\label{eq:LMG_mu1}
\end{equation}
%
%
As already stated in Sec. \ref{sec:BCS}, the correction of order $1/N$ to the entropy then reads
%
%
\begin{equation}
\mathcal{E}^{(1)} =\frac{ \mu^{(1)}}{2} \ln\bigg(\F{\mu^{(0)} +1}{\mu^{(0)} -1} \bigg)\, . 
\label{eq:LMG_correction}
\end{equation}
%
%
We emphasize that this is the exact correction of order $1/N$ whereas in Ref. \cite{Barthel06_2}, only its expansion near $h=1$ was given. 
We display in Fig. \ref{fig:correction_LMG}, a comparison between the numerical correction of order $1/N$ and the analytical expression obtained from Eqs. 
(\ref{eq:LMG_mu1})-(\ref{eq:LMG_correction}). 
%
%
%
%

In the vicinity of the critical point $h=1$, one thus has, at this order:
%
%
\begin{eqnarray}
{\rm e}^\mathcal{E}&=& {\rm e}^{\mathcal{E}^{(0)}+\frac{1}{N}\mathcal{E}^{(1)}} \,,\\
&\sim&[\tau(1-\tau)]^{1/2}\frac{(1-\gamma)^{1/4}}{(h-1)^{1/4}} \bigg[1-\frac{3 (1-\gamma)^{1/2}}{8 N (h-1)^{3/2}} \bigg] \, . \nonumber
\end{eqnarray}
%
%
This suggests that the scaling variable is indeed a function of $\gamma$ but has no dependence on $\tau$. Following Ref. \cite{Barthel06_2}, we reformulate the scaling hypothesis for this model by assuming that in the vicinity of the critical point, a physical observable $\Phi$ can be written as the sum of a regular and a singular contribution, 
%
\begin{equation}
  \Phi_N(h,\gamma)=
  \Phi_N^\mathrm{reg}(h,\gamma)+\Phi_N^\mathrm{sing}(h,\gamma)\, .
\end{equation}
%
where
%
\begin{equation}
\label{eq:phi_sing2}
  \Phi_N^\mathrm{sing}(h\simeq 1,\gamma) \sim   \frac{ (h-1)^{\xi_\Phi^h} (1-\gamma)^{\xi_\Phi^\gamma}} {N^{n_\Phi}}
  \mathcal{G}_\Phi (\zeta)\, ,
\end{equation}
%
and $\zeta=N(h-1)^{3/2}(1-\gamma)^{-1/2}$ is the scaling variable. The exponents $\xi_\Phi^h$, $\xi_\Phi^\gamma$ and $n_\Phi$ are characte\-ristics of the observables $\Phi$. Thus, at the critical point, one expects $\mathcal{G}_\Phi (\zeta) \sim \zeta^{-2 \xi_\Phi^h /3}$, so that 
%
\begin{equation}
\label{eq:phi_sing3}
\Phi_N^\mathrm{sing}(h=1,\gamma)\sim N^{-(n_\Phi + 2 \xi_\Phi^h /3)} (1-\gamma)^{\xi_\Phi^\gamma+ \xi_\Phi^h /3}\, .
\end{equation}
%

We emphasize that the scaling form (\ref{eq:phi_sing2}), can be checked up to high order from the expansion of the ground state energy, the gap, the magnetization, and the spin-spin correlation functions given in Ref. \cite{Dusuel05_2}.

\begin{figure}[t]
  \centering
\includegraphics[width=8cm]{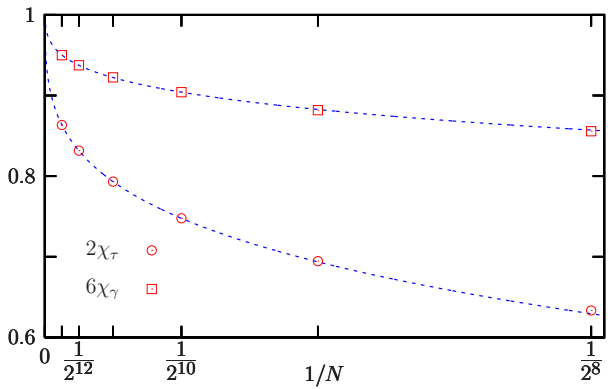}
  \caption{(Color online)
    Exponents $\chi_\gamma$ and $\chi_\tau$ as a function of $1/N$ obtained from numerical diagonalization of $H$. For clarity, we plotted $2 \chi_\tau$ and $6 \chi_\gamma$ which are expected to be equal to 1 in the thermodynamic limit (dotted lines are guides for the eyes).}
  \label{fig:exponents_LMG}
\end{figure}

\begin{figure}[t]
  \centering
\includegraphics[width=8cm]{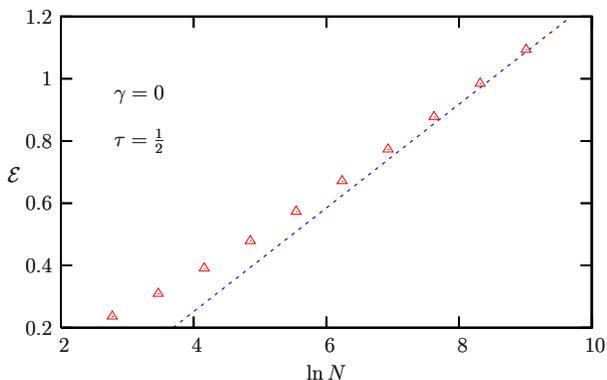}
  \caption{(Color online)
 Entropy as a function of $\ln N$ for $\gamma=0$ and $\tau=1/2$. The blue line has a slope $\chi_N=1/6$.}
  \label{fig:scalingN_LMG}
\end{figure}

The scaling law (\ref{eq:phi_sing3}) applied to the observable ${\rm e}^\mathcal{E}$ leads to the following critical behavior of the entropy 
%
\begin{equation}
\label{eq:entropy_critical}
\mathcal{E}(h=1) \sim \chi_N \ln N+ \chi_\gamma \ln (1-\gamma) + \chi_\tau \ln [\tau(1-\tau)]\, ,
\end{equation}
%
with $\chi_N = \chi_\gamma=1/6$ and  $\chi_\tau =1/2$. As can be seen in Figs. \ref{fig:exponents_LMG} and \ref{fig:scalingN_LMG}, an excellent agreement between these predictions and numerics is observed when increasing the system sizes.
%
%
%
%
%
%
%
%

Such a behavior has striking similarities with the 1D $XY$ model \cite{Latorre03,Latorre04_1,Peschel04,Its05,Its06,Francini06}. 
This is due to the fact that the ground state of the LMG model lies in the maximum spin sector $S=N/2$. As a consequence, the relevant subsystem Hilbert space is spanned by the $L+1$ eigenstates of $\{{\bf (S^\mathcal{A})}^2,S^\mathcal{A}_z\}$ with $S^\mathcal{A}=L/2$. The entanglement entropy as a function of the subsystem size $L$ can thus not exceed the maximum value $\ln(L+1)$. However, note that the physics of the LMG model is very different from 1D models.

This surprising analogy with 1D $XY$ model carries over to the broken phase. Indeed, at order $N^0$, one can compute the entropy in the same way as in the broken phase. Therefore, one first has to perform a rotation around the $y$-axis to bring the $z$-axis along the classical ground state magnetization $\vect{S}^{cl.}=N/2 (\pm\sin\theta_0,0,\cos\theta_0)$ with $\theta_0=\arccos h$ \cite{Dusuel05_2}. To use the HP mapping, we thus need to rotate the spin operators according to
%
%
\begin{equation}
      \label{eq:rotation}
      \left(
        \begin{array}{c}
          S_x\\ S_y\\ S_z
        \end{array}
      \right)
      =
      \left(
        \begin{array}{ccc}
          \cos\theta_0 & 0 & \sin\theta_0\\
          0 & 1 & 0\\
          -\sin\theta_0 & 0 & \cos\theta_0
        \end{array}
      \right)
      \left(
        \begin{array}{c}
          \tS_x\\ \tS_y\\ \tS_z
        \end{array}
      \right)\, ,
\end{equation}
%
%
Note that this choice corresponds to study quantum fluctuations around one of the two classical minima (see \cite{Dusuel05_2} for discussion). Next, one uses the HP representation of $\tS$ which is exactly the same as in Eqs.(\ref{eq:HP1A})-(\ref{eq:HP2B}). The Hamiltonian at order $N^0$ then reads:
%
%
\begin{eqnarray}
  \label{eq:H0}
  H&=&-\frac{N (1+h^2)}{4}-\frac{h^2+\gamma}{4}+\frac{2-h^2-\gamma}{2} \Big( \cra \ana+\crb \anb \Big)  +\nonumber \\
  &&
\frac{\gamma-h^2}{4}  \Big[ \tau \Big( {\cra}^2+\ana^2 \Big) + (1-\tau) \Big( {\crb}^2+\anb^2 \Big) + \nonumber \\
&& 2 \sqrt{\tau(1-\tau)} \Big( \cra \crb +\ana \anb \Big) \Big] + {\mathcal O}(1/N)\, .
\end{eqnarray}
%
%
As can be easily checked, this Hamiltonian coincides with (\ref{eq:HP_LMG}) for $h=1$ as it should for a second order QPT. 
Using the general results of Sec. \ref{sec:general}, one straightforwardly gets the two excitation energies
%
%
\begin{eqnarray}
\Delta_+&=&\frac{2-h^2-\gamma}{2} \, , \\
\Delta_-&=&\sqrt{(1-h^2)(1-\gamma)} \, ,
\end{eqnarray}
%
%
and
%
%
\begin{equation}
\mu^{(0)}=\sqrt{1+\tau(1-\tau) \Bigg[ \bigg(\F{1-h^2}{1-\gamma}\bigg)^{1/4} -
\bigg(\F{1-\gamma}{1-h^2}\bigg)^{1/4} \Bigg]^2}\, .
\end{equation}
%
%
It is worth noting that for $h=\sqrt{\gamma}$, one has $\mu^{(0)}=1$ so that the entropy vanishes (see Fig. \ref{fig:entropy_LMG}).  At this point, the ground state (which is two-fold degenerate in this phase) can indeed be chosen  as separable \cite{Dusuel05_2}. 
Such a point is often present in anisotropic systems \cite{Kurmann82}. \\
%
%
\subsection*{The isotropic case}
%
For $\gamma=1$, the above expression of the gaps are still valid in the thermodynamical limit and one recovers the Goldstone mode predicted by the classical approach. This gapless excitation prevents from using the scaling hypothesis to determine the behavior of the entropy. Fortunately, the isotropic case is trivial since the eigenstates of the LMG Hamiltonian 
(\ref{eq:ham_LMG}) are then the eigenstates $|S,M \rangle$ of   ${\bf S}^2$ and $S_z$. The ground state is obtained for $S=N/2$ and $M<N/2$ for $h<1$,  whereas  for $h>1$ one has $M=N/2$. Thus, in the symmetric phase, the entropy vanishes since the state $|N/2,N/2 \rangle$ is separable, but in the broken phase, the entropy of the Dicke state $|N/2,M<N/2 \rangle$ is known to diverge as $\F{1}{2} \ln N$ \cite{Popkov05,Latorre05_2}. This means that the entropy is discontinuous at the critical point $h=1$ (as the concurrence  \cite{Dusuel05_2}) although this transition is often considered as second-order.
%
%
\section{Conclusion and perspectives}
\label{sec:conclusion}
%
%
The results presented in this paper show that the entanglement entropy in collective models displays a large variety of behaviors. First of all one has to distinguish with respect to the number of Goldstone modes in the system (0, 1 or 2).
%
%
\begin{enumerate}

\item The classical ground state has a finite degeneracy, i.e.\ no Goldstone 
mode: this situation is met in the broken phases of the Dicke and anisotropic 
LMG model, as well as in the symmetric phases of the models studied here. In 
this case, and away from critical points, both excitations around any 
classical states are gapful and the entropy is finite (possibly vanishing). 

\item The classical ground state has one Goldstone mode: this situation is met in the broken phases of the two-level BCS model and the isotropic LMG model. In this case, one excitation energy is finite and the other vanishes and the entropy diverges as  $\F{1}{2} \ln N$. 

\item The classical ground state has two Goldstone modes: this situation is met in the antiferromagnetic Lieb-Mattis model. In this case, the entropy diverges as  $\ln N$.

\end{enumerate}
%
%
In case 1, where we have no Goldstone mode in the system, the collective 
spins, $S^\mathcal{A}$ and  $S^\mathcal{B}$, of the two subsystems may be 
expanded in $1/N$ around their classical orientation via Holstein-Primakoff 
transformation. The system is thus mapped to a two-mode boson model. The 
general analysis of the order $N^0$ (at which the effective bosonic models 
are quadratic) in Sec.\ \ref{sec:general} shows how the entanglement entropy 
diverges when excitation gaps vanish. Its scaling can thus be classified according to the number vanishing of gaps, when a critical point is approached. In the anisotropic LMG model as in the Dicke model, there is only one vanishing gap and the entropy diverges as $\F{1}{6} \ln N$ in both models.

In the two-level BCS model, there are two vanishing gaps and we obtained an entropy that diverges as $\F{1}{3} \ln N$. 
However, these behaviors are found using the scaling hypothesis which relies on the scaling variable $N (g-g_{\rm c})^\alpha$ (see Appendix \ref{App:scaling}). In models studied in this paper,  one always has $\alpha=3/2$ due to their relationship with a quartic oscillator \cite{Leyvraz05}. Nevertheless, this value depends on the Hamiltonian and would be different if, for instance, at the critical point, the terms of order $N^0$ and $1/N$ vanish simultaneously. In this case, the Hamiltonian would be sextic instead of quartic in the creation and annihilation operators and the scaling of the entropy would be different \cite{Emary05}.  
Note also that in the isotropic LMG model, the entropy vanishes at the critical point even though there are two vanishing gaps at the critical point. 

The analysis performed here for simple models corroborates the results obtained in many 1D systems where it has been also found that diverging entropies are always associated to gapless excitations \cite{Holzhey94,Latorre03,Latorre04_1,Korepin04,Peschel04,Its05}. 
This should constitute a strong motivation to further investigate collective models which are simpler to treat than low-dimensional ones and already exhibit nontrivial behaviors of the entanglement entropy. 

%
%
\acknowledgments
%
%
T.~B.\ thanks the DFG and the Studienstiftung des Deutschen Volkes for 
financial support.
%
%
\appendix
%
%

%
%
\section{The scaling hypothesis}
\label{App:scaling}
%
%
This appendix is dedicated to a basic description of the scaling hypothesis. This hypothesis is useful in collective models where a $1/N$ expansion is meaningful such as those considered in this paper. It relies on the fact that, in these models, the $1/N$ expansion of the expectation value of any physical observable 
can be written as the sum of a regular and a singular contribution, 
%
\begin{equation}
  \Phi_N(h,\gamma)=
  \Phi_N^\mathrm{reg}(g)+\Phi_N^\mathrm{sing}(g)\, .
\end{equation}
%
Here, singular means that the function and/or its derivatives with respect to the control parameter $g$ driving the transition diverge at the critical point $g_{\rm c}$, following a power law.
In addition, one has 
%
\begin{equation}
\label{eq:phi_sing1}
  \Phi_N^\mathrm{sing} (g\sim g_{\rm c}) \sim   \F{(g-g_{\rm c})^{\xi_\Phi}}{N^{n_\Phi}}
  \mathcal{F}_\Phi [N (g-g_{\rm c})^{\alpha}]\, .
  \end{equation}
%
The exponents $\xi_\Phi$ and $n_\Phi$ are characteristics of the observables $\Phi$ but the exponent
$\alpha$ depends on the model considered as early understood by Botet and Jullien \cite{Botet83}. 
For the models considered in this paper,  we have explicitly checked that $\alpha=3/2$ which is reminiscent of the behavior of the gap in the vicinity of the critical point which vanishes as $(g-g_{\rm c})^{1/2}$.

The behavior of $\Phi_N^\mathrm{sing}$ at the critical point $g_{\rm c}$ is then obtained by noting that since no divergence can occur at finite $N$, one must have $ \mathcal{F}_\Phi (x), \sim x^{-\xi_\Phi /\alpha}$ and consequently
%
\begin{equation}
\label{eq:phi_sing2}
\Phi_N^\mathrm{sing}(g=g_{\rm c})\sim N^{-(n_\Phi + \xi_\Phi / \alpha)}\, . 
\end{equation}
%

Finally, since we are mainly interested here in the entropy, let us simply note that to obtain a power-law behavior, the observable to be considered is ${\rm e}^{{\mathcal E}}$ which indeed bevaves as suggested in (\ref{eq:phi_sing1}). As a result, if ${\mathcal E} \sim -\xi_{\mathcal E} \ln (g-g_{\rm c})$ near the critical point, the scaling hypothesis predicts that it diverges as  $\F{\xi_{\mathcal E}}{\alpha} \ln N$.

This scaling hypothesis has been checked up to high order in the several models and allowed us to obtain the  finite-size scaling exponents in many systems \cite{Dusuel04_3,Dusuel05_1,Dusuel05_2,Dusuel05_3,Dusuel05_4,Vidal06_2,Barthel06_2}.

%
%
\section{Linear entropy for quadratic two-mode boson systems}
\label{App:entropy_lin}
%
%
In this Appendix, we derive the expression for the linear entropy for the ground state $| \psi \ket$ of a  general quadratic two-mode boson Hamiltonian of the form (\ref{eq:TwoBosonGen-HamVW}). This linear entropy is defined as
%
\begin{equation}
\mathcal{E}_{\rm lin.} =1-\Tr (\rho_{\mathcal{A}}^2)=1-\Tr (\rho_{\mathcal{B}}^2)\, ,
\end{equation}
%
where $\rho_{\mathcal{A},\mathcal{B}}=\mathrm{Tr}_{\mathcal{B},\mathcal{A}} \left(|\psi\rangle\langle\psi|\right)$. Here, we suppose that bosons $a$ and $b$ define subsets $\mathcal{A}$  and $\mathcal{B}$.

As shown in Ref.\cite{Bombelli86},  the reduced density matrices $\rho_{\mathcal{A},\mathcal{B}}$ can always be written as the exponential of a quadratic form in the operators related to the corresponding subset. Concretely, tracing out over $\mathcal{B}$, one has $\rho_{\mathcal{A}}={\rm e}^{-K}$ where 
%
%
\begin{equation}
  K=\kappa_0+\kappa_1 a^\dag a+\kappa_2 \Big( {a^\dag}^2+a^2 \Big)\, .
\end{equation}
%
%
The coefficients $\kappa_i$'s are determined from the conditions
%
%
\begin{equation}
  \label{eq:constraints0}
  \mathrm{Tr}_\mathcal{A}\rho_\mathcal{A}=1,\;\;
  \langle\! \langle a^\dag a \rangle\!\rangle=\langle a^\dag a \rangle
  \mbox{ and }
  \langle\!\langle {a^\dag}^2 \rangle\!\rangle=\langle {a^\dag}^2 \rangle,
\end{equation}
%
%
where $\langle \Omega \rangle= \langle\psi|\Omega|\psi\rangle$ and
$\langle\!\langle \Omega \rangle\!\rangle =\mathrm{Tr}_\mathcal{A}({\rm e}^{-K}\Omega)$. 

The next step is to diagonalize $K$ which becomes 
%
%
\begin{equation}
  K=E_0+\Delta c^\dag c\,,
\end{equation}
%
%
where, as shown in Ref. \cite{Barthel06_1}, 
%
%
\begin{equation}
\Delta= \ln \F{\mu+1}{\mu-1}\, ,
\end{equation}
%
%
and $\mu=\sqrt{-\bra \big(a^\dag+a\big)^2 \ket \bra \big(a^\dag-a \big)^2 \ket}$. 
The normalization condition  $\mathrm{Tr}_\mathcal{A}\rho_\mathcal{A}={\rm e}^{-E_0}/(1-{\rm e}^{-\Delta})=1$ then implies ${\rm e}^{-E_0}=2/(\mu+1)$. It is then straightforward to compute 
%
\begin{eqnarray}
\mathcal{E}_{\rm lin.} &=&1-\Tr {\rm e}^{-2(E_0+\Delta c^\dag c)} \, ,\\
&=&1- \F{1}{\mu}\, .
\end{eqnarray}
%

Note that it is trivial to generalize this derivation for an arbitrary number of bosonic or fermionic modes using the expression of the reduced density matrices given in Ref. \cite{Barthel06_1}.

%
%
\section{First-order correction of the entanglement entropy in a two-mode boson system}
\label{App:correction}
%
%
We give the exact expression of the first-order correction of the entanglement entropy in a two-mode boson system. This correction is derived in Ref. \cite{Barthel06_4} for  the general case.

Let us consider a Hamiltonian of the following form
%
%
\begin{equation}
H=H_0+ g H_1\, ,
\end{equation}
%
%
where $H_0$ is a quadratic two-mode boson Hamiltonian written in terms of bosons $a$ and $b$, and where $g \ll 1$. Using the same notations as in Appendix \ref{App:entropy_lin}, we introduce the parameter 
%
%
\begin{eqnarray}
\mu&=&\sqrt{-\bra \big(a^\dag+a\big)^2 \ket \bra \big(a^\dag-a \big)^2 \ket}\\
&=& \mu^{(0)}+ g \, \mu^{(1)}+\mathcal{O}(g^2) \, ,
\end{eqnarray}
%
%
Then, as detailed in Ref. \cite{Barthel06_4}, the entanglement entropy at order $g^0$ and $g$ is obtained by expanding the function 
%
%
\begin{equation}
F(\mu)=\F{\mu+1}{2}\ln\F{\mu+1}{2}-\F{\mu-1}{2}\ln\F{\mu-1}{2}\, ,
\end{equation}
%
%
at the corresponding order. One then obtains the entanglement entropy between both modes
%
%
\begin{eqnarray}
\mathcal{E}&=&-\mathrm{Tr}_\mathcal{A}\left(\rho_\mathcal{A}\ln\rho_\mathcal{A}\right) \, , \\ 
&=& \mathcal{E}^{(0)}+ g \, \mathcal{E}^{(1)}+\mathcal{O}(g^2) \, ,
\end{eqnarray}
%
%
with 
%
%
\begin{eqnarray}
\mathcal{E}^{(0)}&=&\F{\mu^{(0)}+1}{2}\ln\F{\mu^{(0)}+1}{2}-\F{\mu^{(0)}-1}{2}\ln\F{\mu^{(0)}-1}{2}\, , \quad  \quad \\
\mathcal{E}^{(1)} &=& \frac{ \mu^{(1)}}{2} \ln\bigg(\F{\mu^{(0)} +1}{\mu^{(0)} -1} \bigg)\, . 
\end{eqnarray}
%
%

The only difficulty is thus to compute $\mu$ at order $g$ and one readily has the correction of order $g$ to the entropy. Note finally that the second order correction is \emph{not} obtained by expanding $F$ at order $g^2$.


\end{document}